\documentclass
[reprint,aps,amsmath,amssymb,twoside,prd,showkeys,superscriptaddress]{revtex4-1}
\usepackage[utf8]{inputenc}
\usepackage{multirow}
\usepackage{color}
\usepackage{latexsym}
\usepackage{epsfig,graphics,subfig,float}
\usepackage{epstopdf}

\usepackage{float}
\usepackage{tabularx,booktabs,caption}% http://ctan.org/pkg/{tabularx,booktabs,caption}
\usepackage{amssymb}
\usepackage{verbatim}
\usepackage{lipsum, color}
\usepackage{tikz}
\usepackage{amsmath,amssymb,amsfonts}
\usepackage{tikz}
\usepackage{bm}
\usetikzlibrary{matrix}
\newcommand{\afffias}{Frankfurt Institute for Advanced Studies (FIAS), Ruth-Moufang-Strasse~1, 60438 Frankfurt am Main, Germany}
\newcommand{\affjwg}{Goethe-Universit\"at, Max-von-Laue-Strasse~1, 60438~Frankfurt am Main, Germany}
\newcommand{\affbgu}{Physics Department, Ben-Gurion University of the Negev, Beer-Sheva 84105, Israel}
\newcommand{\affbahamas}{Bahamas Advanced Study Institute and Conferences, 4A Ocean Heights, Hill View Circle, Stella Maris, Long Island, The Bahamas}
\begin{document}

\title{Unified Dark Energy and Dark Matter from Dynamical Space Time}

\author{David Benisty}
\email{benidav@post.bgu.ac.il}
\affiliation{\afffias}\affiliation{\affjwg}\affiliation{\affbgu}
\author{Eduardo I. Guendelman}
\email{guendel@bgu.ac.il}
\affiliation{\afffias}\affiliation{\affbgu}\affiliation{\affbahamas}

\begin{abstract}
A unification of dark matter and dark energy based on a dynamical space time theory is suggested. By introducing a dynamical space time vector field $\chi_\mu$ as a Lagrange multiplier, a conservation of an energy momentum tensor $T^{\mu\nu}_{(\chi)}$ is implemented. This Lagrangian generalizes the "Unified dark energy and dark matter from a scalar field different from quintessence" [Phys.RevD 81, 043520 (2010)] which did not consider a Lagrangian formulation. This generalization allows the solutions which were found previously, but in addition to that also non singular bouncing solutions that rapidly approach to the $\Lambda$CDM model. The dynamical time vector field exactly coincides with the cosmic time for the a $\Lambda$CDM solution and suffers a slight shift (advances slower) with respect to the cosmic time in the region close to the bounce for the bouncing non singular solutions. In addition we introduced some exponential potential which could enter into the $T^{\mu\nu}_{(\chi)}$ stress energy tensor or coupled directly to the measure $\sqrt{-g}$, gives a possible interaction between DE and DM and could explain the coincidence problem.
\end{abstract}

\maketitle

\section{Introduction}
Dark energy and Dark matter constitute most of the observable Universe. Yet the true nature of these two phenomena is still a mystery. One fundamental question with respect to those phenomena is the coincidence problem which is trying to explain the relation between dark energy and dark matter densities. In order to solve this problem, one approach claims that the dark energy is
a dynamical entity and hope to exploit solutions of scaling or tracking type to remove dependence on initial conditions. Others left this principle and tried to model the dark energy as a phenomenological fluid which exhibits a particular relation with the scale factor \cite{Cardone:2004sq}, Hubble constant \cite{Dvali:2003rk} or even even the cosmic time itself \cite{Basilakos:2009ah}. 

Unifications between dark energy and dark matter from an action principle were obtained from K-essence type actions \cite{Scherrer:2004au}, or by introducing a complex scalar field \cite{Arbey:2006it}. Beyond those approaches, a unified description of Dark Energy and Dark Matter using a new measure of integration has been formulated \cite{Guendelman:2015jii}-\cite{Guendelman:2016kwj} . Also a diffusive interacting of dark energy and dark matter models was introduced in \cite{Koutsoumbas:2017fxp}\cite{Haba:2016swv} and it has been found that  diffusive interacting dark energy - dark matter models can be formulated in the context of an action principle based on a generalization of those Two Measures Theories in the context of quintessential scalar fields \cite{Benisty:2017eqh}\cite{Benisty:2017rbw}, although these models are not equivalent to the previous diffusive interacting dark energy - dark matter models \cite{Koutsoumbas:2017fxp}\cite{Haba:2016swv}. 

One has to take now into consideration the measurements in 17 August of 2017 of multi-messenger gravitational wave astronomy which are in contradiction to many modified theories of gravity predictions. These observations have commenced with the detection of the binary neutron star merger GW170817 and its associated electromagnetic counterparts \cite{TheLIGOScientific:2017qsa}. Both signals place an exquisite bound on the speed of gravity to be  the same as the speed of light. This constraint rejected many modifications to GR \cite{Creminelli:2017sry}-\cite{Lombriser:2016yzn} and also many unifications between dark energy and dark matter. 

A model, which also continues to be valid after GW170817 event, for a unification of dark energy and dark matter from a single scalar field $\phi$, was suggested by Gao, Kunz, Liddle and Parkinson \cite{Gao:2009me}. Their model is close to traditional quintessence, and gives dynamical dark energy and dark matter, but introduces a modification of the equations of motion of the scalar field that apparently are impossible to formulate in the framework of an action principle. The basic stress energy tensor which was considered in addition to Einstein equation was:
\begin{equation}\label{original}
T^{\mu\nu} = -\frac{1}{2} \phi^{,\mu} \phi^{,\nu} + U(\phi) g^{\mu \nu}  
\end{equation}
where $\phi$ is a scalar field and $U(\phi)$ is the potential for that scalar. Assuming homogeneous and isotropic behavior the scalar field should be only time dependent $\phi=\phi(t)$. Then the kinetic term $-\frac{1}{2} \phi^{,\mu} \phi^{,\nu}$ is parameterzing the dark matter because  it contains only energy density with no pressure and $U(\phi)g^{\mu\nu}$ is parameterzing the dark energy. The basic requirement for this stress energy tensor is it's conservation law $\nabla_\mu T^{\mu\nu}=0$. By assuming a constant potential $U(\phi) = \textbf{Const}$ the model provides from the potential the traditional cosmological constant and the kinetic term of the scalar field is shown to provide, from the conservation law of the energy momentum tensor, that the kinetic term dependence has a dust like behavior.
\begin{equation}
-\frac{1}{2}\nabla_\mu (\phi^{,\mu} \phi^{,\nu}) = 0 \quad  \Rightarrow \quad \dot{\phi}^2 \sim \frac{1}{a^3}
\end{equation}
This simple case refers to the classical $\Lambda$CDM model. The special advantage of this model is a unification of dark energy and dark matter from one scalar field and has an interesting possibility for exploring the coincidence problem.

The lack of an action principle for this model brought us to reformulate the unification between dark energy and dark matter idea put forward by  Gao, Kunz, Liddle and Parkinson \cite{Gao:2009me} in the framework of a Dynamical Space Time Theory \cite{Guendelman:2009ck}\cite{Benisty:2016ybt} which forces a conservation of energy momentum tensor in addition to the covariant conservation of the stress energy momentum tensor that appears in Einstein equation. In the next chapter we explore the equations of motion for these theories. In the third chapter we solve analytically the theory for constant potentials which reproduce the $\Lambda$CDM model with a bounce, which gives a possibility to solve the initial big bang singularity. In the last chapter we solve the theory for an exponential potential which gives a good possibility for solving the coincidence problem.

\section{Dynamical Space Time Theory}
\subsection{A basic formulation}
One from the basic fatures in the standard approach to theories of gravity is the local conservation of an energy momentum tensor. In the field theory case it's  derived as a result rather than a starting point. For example, the conservation of energy can be derived from the time translation invariance principle. The local conservation of an energy momentum tensor can be a starting point rather than a derived result. Let's consider a 4 dimensional case where a conservation of a symmetric energy momentum tensor $T^{\mu\nu}_{(\chi)}$ is imposed by introducing the term in the action:
\begin{equation} \label{action}
	\mathcal{S}_{(\chi)}=\int d^{4}x\sqrt{-g}\chi_{\mu;\nu}T_{\left(\chi\right)}^{\mu\nu}
\end{equation}
where $ \chi_{\mu;\nu}=\partial_{\nu}\chi_{\mu}-\Gamma_{\mu\nu}^{\lambda}\chi_{\lambda}$. The vector field $\chi_\mu$ called a dynamical space time vector, because of the energy
density of $T^{\mu\nu}_{(\chi)}$ is a canonically conjugated variable to $\chi_0$, which is what we expected from a dynamical time:
\begin{equation} 
	\pi_{\chi_{0}} = \frac{\partial \mathcal{L}}{\partial \dot{\chi}^0} = T^{0}_{0} (\chi) 
\end{equation}

If $ T_{\left(\chi\right)}^{\mu\nu} $ is being independent of $\chi_{\mu}$  and having $ \Gamma_{\mu\nu}^{\lambda} $ being defined as the Christoffel connection coefficients (the second order Formalism), then the variation with respect to $ \chi_{\mu} $ gives a covariant conservation law:
\begin{equation}
\nabla_{\mu}T_{\left(\chi\right)}^{\mu\nu}=0
\end{equation}
From the variation of the action with respect to the metric, we get a conserved stress energy tensor $G^{\mu\nu}$ (in appropriate units), which is well known from Einstein equation:
\begin{equation}
	G^{\mu\nu}=\frac{2}{\sqrt{-g}}\frac{\delta\sqrt{-g}}{\delta g^{\mu\nu}}[\mathcal{L}_{\chi}+\mathcal{L}_{m}]\,,\quad \nabla_{\mu} G^{\mu\nu}=0\,.
\end{equation}
where $G^{\mu\nu}$ is Einstein tensor, $\mathcal{L}_{\chi}$ is the Lagrangian in (\ref{action}) and $\mathcal{L}_{m}$ is an optional action that involve other contributions. 

Some basic symmetries that holds for the dynamical space time theory are two independent shift symmetries:
\begin{equation}\label{shift}
\chi_\mu \rightarrow \chi_\mu + k_\mu \quad ,\quad T_{\left(\chi\right)}^{\mu\nu} \rightarrow T_{\left(\chi\right)}^{\mu\nu} + \Lambda g^{\mu\nu}
\end{equation}
where $\Lambda$ is some arbitrary constant and $k_\mu$ is a Killing vector of the solution.
This transformation will not change the equations of motions, which means also that the process of redefinition of the energy momentum tensor in the action (\ref{action}) will not change the equations of motion. Of course such type of redefinition of the energy momentum tensor is exactly what is done in the process of normal ordering in Quantum Field Theory for instance.

\subsection{A connection to modified measures}
A particular case of the stress energy tensor with the form $T_{(\chi)}^{\mu\nu}=\mathcal{L}_{1}g^{\mu\nu}$ corresponds to a modified measure theory. By substituting this stress energy tensor into the action itself, the determinant of the metric is canceled:
\begin{equation}\label{mm}
\sqrt{-g}\chi^{\mu}_{;\mu} \mathcal{L}_{1} 
= \partial_\mu(\sqrt{-g}\chi^{\mu}) \mathcal{L}_{1} = \Phi \mathcal{L}_{1}
\end{equation}
where $\Phi=\partial_\mu(\sqrt{-g}\chi^{\mu})$ is like a "modified measure". A variation with respect to the dynamical time vector field will give a constraint on $\mathcal{L}_1$ to be a constant:
\begin{equation}\label{constraint}
{\partial_{\alpha}\mathcal{L}_{1}=0} \quad \Rightarrow \quad \mathcal{L}_{1}=M=Const
\end{equation}
This situation corresponds to the "Two Measures Theory" \cite{TMT1}-\cite{TMT3} where in addition to the regular measure of integration in the action $ \sqrt{-g} $ includes another measure of integration which is also a density and a total derivative. Notable effects that can be obtained in this way are the spontaneous breaking of the scale invariance, the see saw cosmological effects \cite{TMT1}, the resolution of the 5th force problem in quintessential cosmology \cite{TMT4} and a unified picture of both inflation and of slowly accelerated expansion of the present universe \cite{Guendelman:2002js}\cite{Guendelman:2014bva}. As we mentioned before in the introduction Two Measure Theory can serve to build unified models of dark energy and dark matter. 

Usually the construction of this measure is from 4 scalar fields $ \varphi_{a} $, where $ a=1,2,3,4 $. 
\begin{equation}
\Phi=\frac{1}{4!}\varepsilon^{\alpha\beta\gamma\delta}\varepsilon_{abcd}\partial_{\alpha}\varphi^{(a)}\partial_{\beta}\varphi^{(b)}\partial_{\gamma}\varphi^{(c)}\partial_{\delta}\varphi^{(d)}
\end{equation}  
and then we can rewrite an action that uses both of these densities:
\begin{equation}
        S=\int d^{4}x\Phi\mathcal{L}_{1}+\int d^{4}x\sqrt{-g}\mathcal{L}_{2}\,.
\end{equation}

As a consequence of the variation with respect to the scalar fields $ \varphi_{a} $, assuming that $ \mathcal{L}_{1} $ and $ \mathcal{L}_{2} $ are independent of the scalar fields $\varphi_{a}$, we obtain that for $ \Phi\neq0 $ it implies that $\mathcal{L}_{1}=M=const$ as in the dynamical time theory with the case of (\ref{constraint}).

\section{DE-DM Unified Theory from Dynamical Space-Time}
A suggestion of an  action which can produce DE-DM unification takes the form:
\begin{equation}
\mathcal{L}=-\frac{1}{2}R+\chi_{\mu;\nu} T^{\mu\nu}_{(\chi)} - \frac{1}{2}g^{\alpha\beta} \phi_{,\alpha}\phi_{,\beta} - V(\phi)
\end{equation}
Consisting of  an Einstein Hilbert action ($8 \pi G=1$), quintessence and Dynamical space-time action, when the original stress energy tensor $T^{\mu\nu}_{(\chi)}$ is the same as the stress energy tensor (\ref{original}) Gao and colleagues used:
\begin{equation}\label{tmunuchi}
T^{\mu\nu}_{(\chi)} = -\frac{1}{2} \phi^{,\mu} \phi^{,\nu} + U(\phi) g^{\mu \nu}  
\end{equation}
The action depends on three different variables: the scalar field $\phi$, the dynamical space time vector $\chi_\mu$ and the metric $g_{\mu\nu}$. Therefore there are 3 sets in for the equation of motions. For the solution we assume homogeneity and isotropy, therefore we solve our theory with a FLRW metric:
\begin{equation}
ds^2=-dt^2+a(t)^2(\frac{dr^2}{1-K r^2}+r^2 d\Omega^2)
\end{equation}
According to this ansatz, the scalar field is just a function of time $\phi(t)$ and the dynamical vector field will be taken only with a time component $\chi_\mu = (\chi_0,0,0,0)$, where $\chi_0$ is also just a function of time.
A variation with respect to the dynamical space time vector field $\chi_\mu$ will force a conservation of the original stress energy tensor, which in FRWM gives the relation:
\begin{equation}\label{1frw}
\ddot{\phi}+\frac{3}{2}\mathcal{H}\dot{\phi}+U'(\phi)=0
\end{equation}
Compared with the equivalent equation which comes from quintessence model, this model gives a different and smaller friction term, as compared to the canonical scalar field. Therefore for increasing redshift, the densities for the scalar field will increase slower than in the standard quintessence. 

The second variation with respect to the scalar field $\phi$ gives a non-conserved current: 
\begin{subequations}
\begin{equation}
\chi^\lambda_{;\lambda} U'(\phi) - V'(\phi) = \nabla_\mu j^{\mu} 
\end{equation}
\begin{equation}
j^{\mu}  = \frac{1}{2}\phi_{,\nu} (\chi^{\mu;\nu}+\chi^{\nu;\mu}) + \phi^{,\mu}
\end{equation}
\end{subequations}
and the derivatives of the potentials are the source of this current. For constant potentials the source becomes zero, and we get a covariant conservation of this current. In a FLRW metric this equation of motion takes the form:
\begin{widetext}
\begin{equation}\label{15}
\ddot{\phi} (\dot{\chi}_0 - 1) + \dot{\phi} [\ddot{\chi}_0+3 \mathcal{H}(\dot{\chi}_0 - 1)] =  U'(\phi) (\dot{\chi}_0 + 3\mathcal{H}\chi_0) - V'(\phi)
\end{equation}
Substituting the term of the potential derivative $U'(\phi)$ from equation (\ref{1frw}):  
\begin{equation}\label{chi}
[1-2\dot{\chi}_0-3\mathcal{H}\chi_0]\ddot{\phi}-[\ddot{\chi}_0-3\mathcal{H}+\frac{9}{2}\mathcal{H}(\dot{\chi}_0+\chi_0\mathcal{H})] \dot{\phi}+V'(\phi)=0
\end{equation}
The last variation, with respect to the metric, gives the stress energy tensor that is defined by the value of the Einstein's tensor:
\begin{equation}
G^{\mu\nu}= g^{\mu\nu} (\frac{1}{2}\phi_{,\alpha} \phi^{,\alpha} +  V(\phi)
+\frac{1}{2}\chi^{\alpha;\beta}\phi_{,\alpha} \phi_{,\beta}+\chi^{\lambda}\phi_{,\lambda}U'(\phi)) 
- \frac{1}{2} \phi^{,\mu} ((\chi^{\lambda}_{;\lambda}+2) \phi^{,\nu} + \chi^{\lambda;\nu}\phi_{,\lambda} + \chi^\lambda \phi^{,\nu}_{;\lambda})
- \frac{1}{2}(\chi^{\lambda}\phi^{,\mu}_{;\lambda}\phi^{,\nu}+\chi^{\lambda;\mu}\phi_{,\lambda}\phi^{,\nu})
\end{equation}
\end{widetext}
For the spacially homogeneous, cosmological case, the energy density and the pressure of the scalar field are:
\begin{subequations}
\begin{equation}
\rho=\dot{\phi}^2 (\dot{\chi}_0(1-\frac{3}{2}\mathcal{H})-\frac{1}{2}) + V(\phi) -\dot{\phi}\dot{\chi}_0 (U'(\phi)+\ddot{\phi})
\end{equation}
\begin{equation}
p = \frac{1}{2}\dot{\phi}^2(\dot{\chi}_0-1) - V(\phi) -\chi_0 \dot{\phi} U'(\phi)
\end{equation}
\end{subequations}
\begin{figure}[h]
 	\centering
\includegraphics[width=0.5\textwidth]{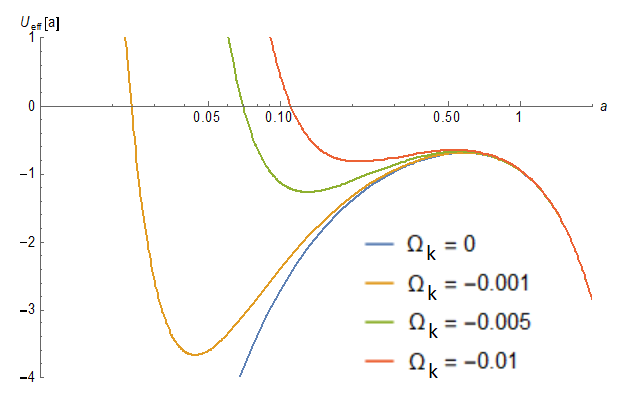}
\caption{Plot of the effective potential. For $\Omega_\kappa \neq 0 $, there is a bouncing universe with dynamical dark energy.}
 	\label{fig1}
 \end{figure}
Substituting the potential derivative $U'(\phi)$ from Eq. (\ref{1frw}) into the energy density term, makes the equation simpler: 
\begin{equation}\label{density}
\rho = (\dot{\chi}_0-\frac{1}{2})\dot{\phi}^2 + V(\phi)
\end{equation}
which has no longer dependence on the potential $U(\phi)$ or it's derivatives. Those three variations are sufficient for building a complete solution for the theory. Let's see a few simple cases.

\section{The evolution of the homogeneous solutions}
\subsection{A bouncing $\Lambda$CDM solution}
In order to compute the evolution of the scalar field and to check whether it is compatible with observable universe, we have to specify a form for the potentials. Let's take a simplified case of constant potentials:
\begin{equation}
U(\phi)= C ,\quad  V(\phi)=\Omega_\Lambda
\end{equation}
Overall, in the equations of motions only the derivative the potential $U(\phi)$ appears, not the potential itself. Therefore a constant part of the potential $U(\phi)$ does not contribute to the solution. However $V(\phi)$, as we shall see below, gives the cosmological constant. The conservation of the stress energy tensor from equation (\ref{1frw}) gives:
\begin{equation}\label{phi}
\dot{\phi}^2 = \frac{2\Omega_m}{a^3}
\end{equation}
where $\Omega_m$ is an integration constant which appears from the solution. 
From the second variation, with respect to the scalar field $\phi$, a conserved current is obtained, which from equation (\ref{chi}) gives the exact solution of the dynamical time vector field:
\begin{equation}\label{chidot}
\dot{\chi}_0 = 1 - \kappa\, a^{-1.5}
\end{equation}
where $\kappa$ is another integration of constant. Eventually, the densities and the pressure for this potentials are given by (\ref{density}). By substituting the solutions for the scalar $\dot{\phi}$ and the vector $\dot{\chi}_0$ (in units with $\rho_c = \frac{8\pi G}{3H^2_0}=1$) we get:
\begin{subequations}
\begin{equation}
\rho = \Omega_\Lambda - \frac{\Omega_\kappa}{a^{4.5}} +\frac{\Omega_m}{a^3} 
\end{equation}\begin{equation}
p = -\Omega_\Lambda - \frac{1}{2}\frac{\Omega_\kappa}{a^{4.5}}
\end{equation}
\end{subequations}
 \begin{figure}[h]
 	\centering
\includegraphics[width=0.5\textwidth]{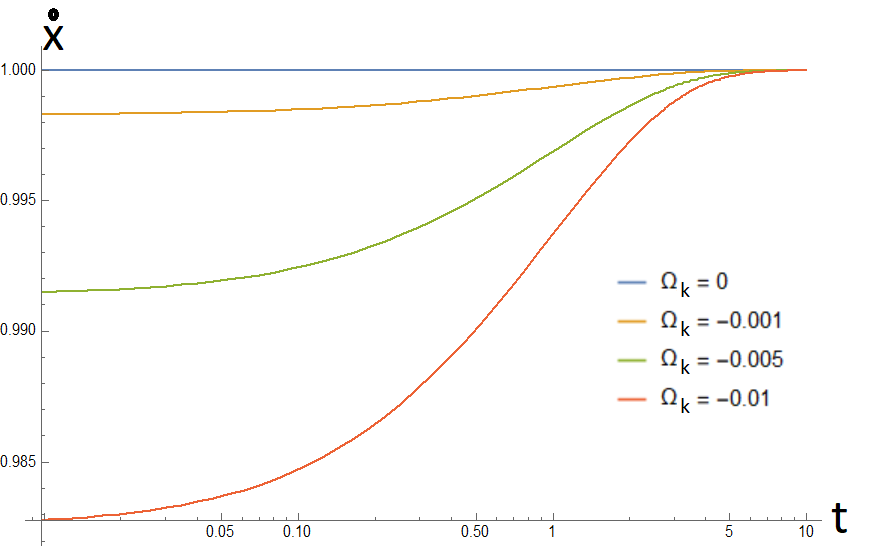}
\caption{Plot of $\dot{\chi}_0$ vs. the cosmic time.}
 	\label{fig2}
 \end{figure} 
  \begin{figure}[b]
 	\centering
 \includegraphics[width=0.5\textwidth]{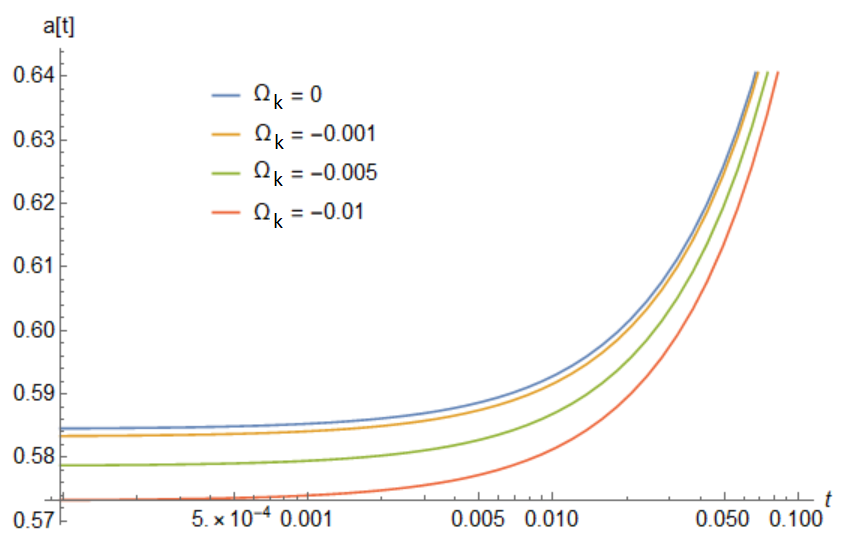}
\caption{Plot of the scale parameter vs. the cosmic time. In any case $\dot\chi_0 \approx 1$} 
\label{fig3}
 \end{figure} 
where $\Omega_\kappa=\kappa\Omega_m$. Notice that $\Omega_m,\Omega_\kappa$ are integration constants the solution contains and $\Omega_\Lambda$ is parameter from the action of the theory. We can separate the result into three different "dark fluids": dark energy ($\omega = -1$), dark matter ($\omega = 0$) and an exotic part ($\omega = \frac{1}{2}$), which is the responsible for the bounce (for $\kappa > 0$). From Eq. (\ref{phi}) the solution produces a positive $\Omega_m$ since it's proportional to $\dot{\phi}^2$. For $\Omega_\Lambda$ the measurements for the late universe forces the choice of this parameter to be positive. However for another solutions (in the context of Anti de-Sitter space, for instance) this parameter could be negative from the beginning.

In Fig (\ref{fig1}) we can see the effective potential for different values of $\Omega_\kappa$. For $\Omega_\kappa = 0$ the solution returns to the known $\Lambda$CDM model. However for $\Omega_\kappa < 0$ we obtain a bouncing solution which also returns to the $\Lambda$CDM for late time expansion. 

In addition to those solutions, there is a strong correspondence between the zero component of the dynamical space time vector field and the cosmic time. For $\Lambda$CDM there is no bouncing solution $\kappa=0$ and therefore from equation (\ref{chidot}) we get $\chi_0=t$ that implies that the dynamical time is exactly the cosmic time. For bouncing $\Lambda$CDM (see Fig.(\ref{fig2})) we obtain a relation between the dynamical and the cosmic time with some delay between the dynamical time and the cosmic time for the early universe (in the bouncing region). For the late universe the dynamical time returns back to run as fast as the cosmic time again. This relation between the dynamical and the cosmic time may have interesting application in the solution to "the problem of time" in quantum cosmology which will discussed elsewhere. Notice that the dynamical time is a field variable while the cosmic time is a coordinate.
\begin{figure*}[t]
 	\centering
 	\includegraphics[width=0.8\textwidth]{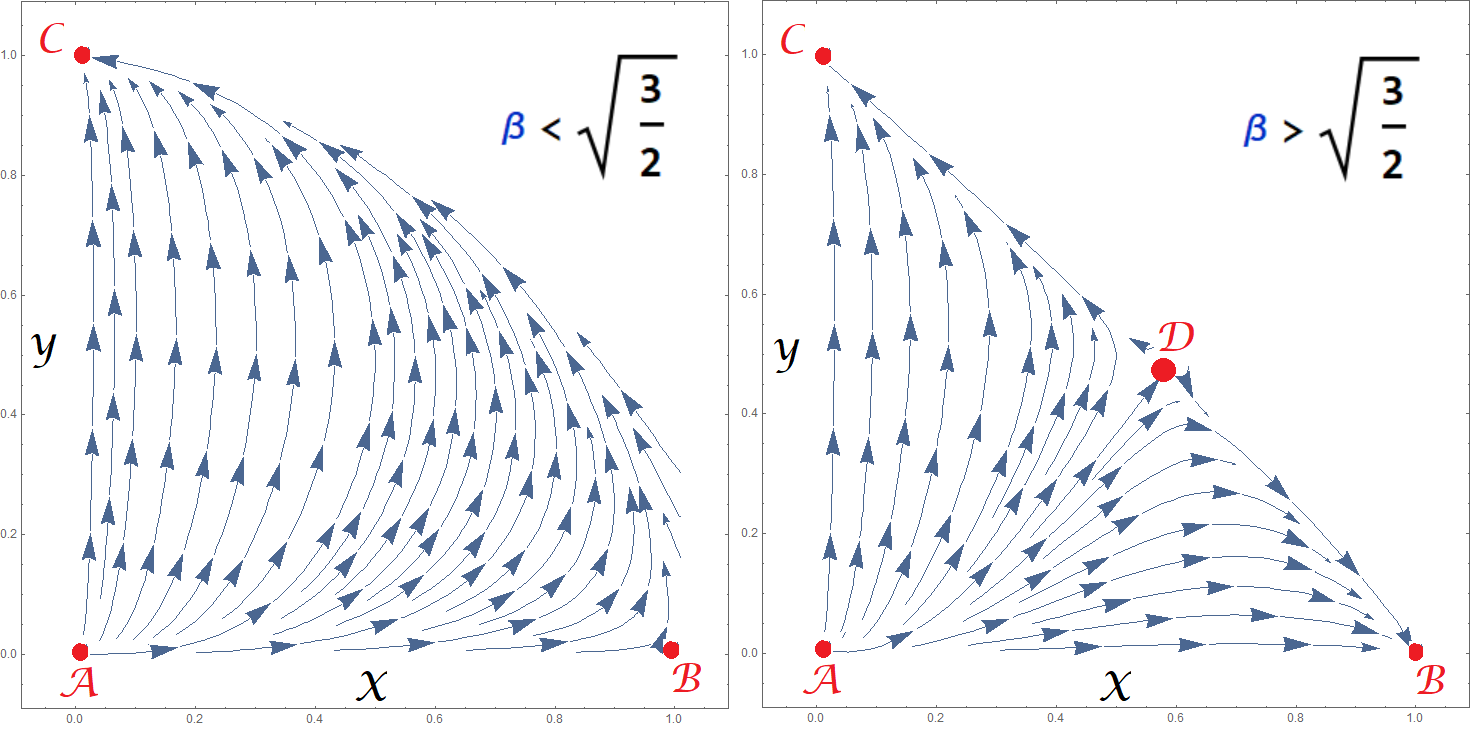}
\caption{The phase plane for different values of $\beta$} 
\label{fig4}
\end{figure*} 
The scale parameter evolution depicted in Fig.(\ref{fig3}) can show us the initial conditions where $\dot{a}(t)=0$, because at that point $a(t)$is a minimum.  In addition, for all cases the initial condition for the scale parameter is not zero $a(0) \neq 0$. These features imply a bouncing universe solutions.
 
\subsection{Interacting DE-DM}
\subsubsection{Autonomous system method}
For studying the evolution of the scalar field in the case of interacting DE-DM we address more generic potentials. For instance:
\begin{equation}
U(\phi)= C ,\quad  V(\phi)=\Omega_\Lambda e^{-\beta \phi}
\end{equation}
where $\beta > 0$ (if not we can perform the transformation $\phi \rightarrow -\phi$). In the limit $\beta \rightarrow 0$ the solution returns to the constant potentials case and therefore the model is continuously connected to $\Lambda$CDM, at least as far as the background evolution is concerned.
The first equation of motion (\ref{1frw}) gives us last case (\ref{phi}), or in this form:
\begin{equation}
\ddot{\phi}=-\frac{3}{2} \mathcal{H}\dot{\phi}
\end{equation}
The equation of motion with respect to the scalar field $\phi$ can be expressed with a new dimensionless parameter:
\begin{equation}
\delta = \dot{\chi}_0-1
\end{equation}
which represents the difference of the rates of change  between the zero component of the dynamical space time vector and the the actual cosmic time. The equation of motion (\ref{15}) in terms of this variable  gets the form: 
\begin{equation}
\dot{\phi}(\dot{\delta}+\frac{3}{2}\mathcal{H}\delta)=\beta V(\phi)
\end{equation}
Notice that for $\beta = 0$ the relation for $\delta = 2\kappa\, a^{-1.5}$ as equation (\ref{chidot}). The main equation of the dynamical system are given by the following dimensionless quantities:
\begin{equation}\label{def}
x = \frac{\dot{\phi}}{\sqrt{6}H} , \quad y = \frac{\sqrt{V(\phi)}}{\sqrt{3}H}
\end{equation}
where $x$ and $y$ are represent the density parameters of the kinetic (dark matter like) and potential (dark energy like) terms, respectively. With those new three parameters $(x,y,\delta)$, the equation of motion with respect to the metric is written as:  
\begin{equation}\label{28}
(1+2\delta)x^2+y^2=1
\end{equation}

Assuming low values of $\beta$ the dynamical time and the cosmic time approximately coincide  (see figure \ref{fig3}) and therefore $\delta \approx 0$. The phase portrait in that case should not deviate too much from a closed circle. Hence, equation (\ref{def}) can be written by the following autonomous system equations:
\begin{subequations}\label{DSM}
\begin{equation}
\frac{dx}{d\tau} = -\frac{3x}{4} (x^2 - 1 + 3 y^2)
\end{equation}

\begin{equation}
\frac{dy}{d\tau} = -\frac{y}{4} (-9 + 3 x^2 + 9 y^2 + 2 \sqrt{6} x \beta)
\end{equation}
\end{subequations}
where $\tau = \ln{a}$. The equation of state $\omega$ also can be written as:

\begin{equation}
\omega_\chi = \frac{1}{2} \left(1-x^2-3 y^2\right)
\end{equation}

\begin{table}[t]
  \begin{center}
  \label{tab:table1}
    \begin{tabular}{c|c|c|c} % <-- Alignments: 1st column left, 2nd middle and 3rd right, with vertical lines in between   
    \textbf{Name} & \textbf{existence} & \textbf{stability} & \textbf{universe}\\
       \hline
      A & all $\beta$ & unstable & - \\
      B &  all $\beta$  & stable for $\beta>\sqrt{\frac{3}{2}}$ & Dark Matter\\
      C & all $\beta$   & asymptotically stable & Dark Energy\\
      D & $\beta>\sqrt{\frac{3}{2}}$ & unstable saddle p. & unified DE-DM\\
    \end{tabular}
    \caption{The properties of the critical points for the exponential potential}
  \end{center}
\end{table}
The properties of a few fixed points for the exponential potential presenting in Table I. The features of the fixed points can separate to two cases.

\begin{figure*}[t!]
    \centering
    \includegraphics[width=0.9\textwidth]{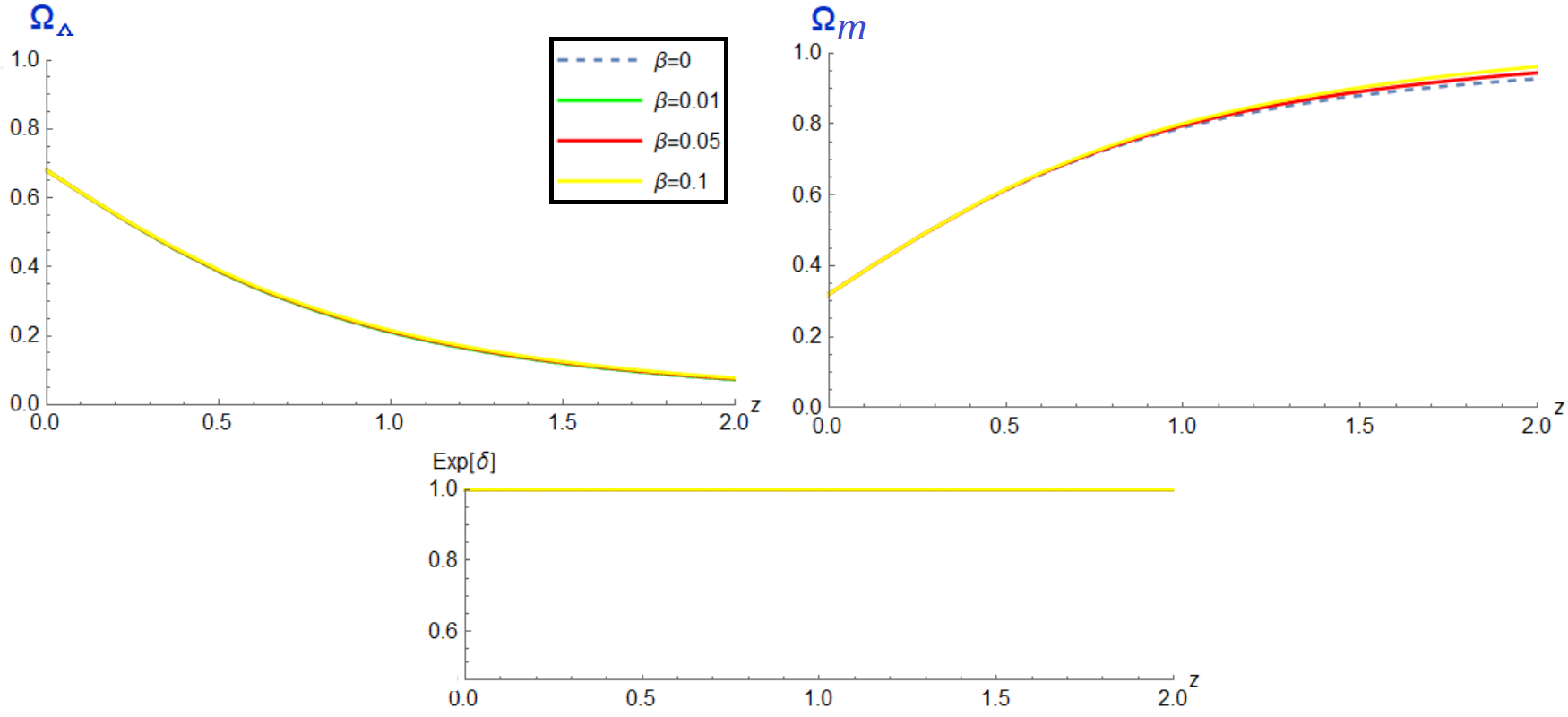}
    \caption{The evolution of DE-DM ratios and $e^\delta \sim 1$ for small values of $\beta$.}
\label{fig5}
 \end{figure*}

One case is when $\beta<\sqrt{\frac{3}{2}}$ and all of the solutions are flowing into a Dark Energy dominated universe (point C $(x=0,y=1)$). The dark matter dominated universe is an unstable point (point B $(x=1,y=0)$) that the universe goes though which corresponds to the dark matter epoch. In any case point A ($(x=0,y=0)$) which representing no dark matter and no dark energy, does not really exist, because of the contradiction to equation (\ref{28}). However if the initial condition starts close to this point it's driven into dark energy dominance eventually, as you can see in figure (\ref{fig5}). 
Also for this case the shape of the phase portrait looks as a circle, which ensure our assumption about the identification between the dynamical space time and the cosmic time. 

In the second case $\beta>\sqrt{\frac{3}{2}}$ and there are two stable fixed point. One for dark energy (C), and one for dark matter (B). If the initial conditions are close enough to those points, it will be attracted into them. In addition, a saddle point D ($x=\sqrt{\frac{3}{2}}\frac{1}{\beta},y=\frac{\sqrt{2 \beta ^2-3}}{\sqrt{6} \beta }$) is obtained. For this point the ratio between the pressure and the density is $\omega=-\frac{2}{3}+\frac{1}{\beta^2}$. Some solutions are attracted to this point, but eventually they are repelled to the closer point. However, the case of $\beta>\sqrt{\frac{3}{2}}$ contradicts the assumption that $\beta$ is small enough in order not to deviate from $\Lambda$CDM.
Also for this case the shape of the phase portrait in this case, deviates from a circle which implies a big deviation between the dynamical space time and the cosmic time.

This modification, which adds one exponential potential, is not the most general case since we could also add an additional potential which would enter into the $T_{(\chi)}^{\mu\nu}$. We suspect that some form of the potentials which are more general, could cause point D becomes stable and will lead us to a more comprehensive understanding of the cosmic coincidence problem, which will investigate in the future. 
\subsubsection{Evolution of Physical Quantities}
In order to assess the viability of the model, lets see how some physical quantities change vs. the redshift ($z$). The connection between the cosmic time derivative and a redshift derivative is:
\begin{equation}
\frac{d}{dt} = - \mathcal{H}(z) (z+1) \frac{d}{dz}
\end{equation}
which has been obtained from the the scale factor dependence on $z$, $a(z) = \frac{a_0}{z+1}$.Figure (\ref{fig5}) describes the cosmological energies densities $\Omega_m,\Omega_\Lambda$ vs. the redshift. For $\beta = 0$ case, which refers to $\Lambda$CDM model (any time we can set $\Omega_\kappa$ to be zero or small) we can see that in earlier times $\Omega_m$ becomes dominant, for earlier times that is for the very early universe, $\Omega_\kappa$ (which we have taken to be very small except for the very early universe) dominates. For different values of $\beta$ we can see a slight shift from $\Lambda$CDM,  which should be more dominant in the early universe. The variable $\delta$, that measures the difference in the evolution of the dynamical time and the cosmic time, which in the case of $\beta = 0$ gives a contribution that can be parametrized by $\Omega_\kappa$, has been taken to be very close to zero in all cases except for the very early universe, because of there a strong impact exists,  close the  bounce that replaces now the traditional big bang.

\begin{figure}[b]
 	\centering
 	\includegraphics[width=0.5\textwidth]{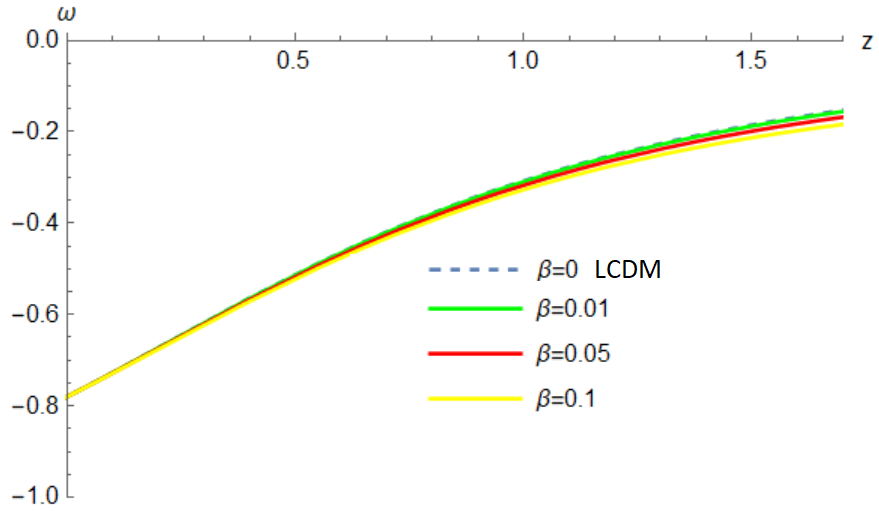}
\caption{The equation of state of the universe for different values of $\beta$.}
 	\label{fig6}
 \end{figure}

In figure (\ref{fig6}) we can see the evolution of the equation of state of whole universe as a function of redshift. It behaves as  cold dark matter dominated at higher redshifts and dark energy for the lower redshifts. The behavior does not tremendously change for those values of the redshift, but the deviations are measurable. 

The set of potentials that where suggested in this chapter have a nice feature which reduce the dependence of the number of quantities. In this way a suggestive and convenient parametrization of the solution uses variable $\delta = \dot{\chi}_0 - 1$ which contains all the dependence on $\chi_0$ . In the future it would be interesting to investigate how different potentials would affect on the physical quantities of the universe. However, unlike other models of dark energy and dark matter, even a trivial assumption of constant potentials leads  directly to a unification of dark energy and dark matter. In any case, any generalization should assume a constant potentials asymptotically.

\section{Discussion and Future work}
In this paper the "unified dark energy and dark matter from a scalar field different from quintessence" is formulated through an action principle. Introducing the coupling of a dynamical space time vector field to an energy momentum tensor that appears in the action, determines the equation of motion of the scalar field from the variation of the dynamical space time vector field or effectively
from the conservation law of an energy momentum tensor, as in \cite{Gao:2009me}. The energy momentum tensor that is introduced in the action is related but not in general the same as the one that appears in the right hand side of the gravitational equations, as opposed to the non Lagrangian approach of
\cite{Gao:2009me}, so our approach and that of \cite{Gao:2009me} are not equivalent. However in many situations the solutions studied in \cite{Gao:2009me} can be also obtained here, but there are other solutions, in special non singular bounce solutions which are  not present in \cite{Gao:2009me}. 

In those simple solutions, the dynamical time behaves very close to the cosmic time. In particular in solutions which are exactly $\Lambda$CDM, the cosmic time and the dynamical time exactly coincide with each other. If there is a bounce, the deviation of the dynamical time with respect to the cosmic time takes place only very close to the bounce region. The use of this dynamical time as the time in the Wheeler de-Witt equation should also a subject of interest.

In principle we can introduce two different scalar potentials: one coupled directly to $\sqrt{-g}$ and the other appearing in the original stress energy tensor $T^{\mu\nu}_{(\chi)}$. So far, for the purposes of starting the study of the theory, we have only introduced a scalar potential coupled directly to $\sqrt{-g}$ and shown that this already leads to an interacting Dark Energy - Dark Matter model, although the full possibilities of the theory will be revealed when the two independent potentials will be introduced.

Possible signatures for this model or for more generalized forms could be could be identified from cosmological perturbations theory. For instance, the perturbation for the scalar field is clear. However, the perturbation for the vector field could be represented with more degrees of freedom which can reproduce a different power spectrum for the Cosmic Microwave Background Anisotropies for instance. But more over than this, the model that was suggested in the last part was only with an exponential potential. However many combinations of potentials are applicable for testing the evolution for the energy densities, and using data fitting for those models. The benefits for this models are that they still preserve the speed of gravity equal to the speed of light, and also that arises from an action principle. Researching those families of solutions with more general potentials could help solve the coincidence problem. 

The effects studied in the context bouncing solution, which can
prevent the initial big bang singularity, could have a consequences for
the radially falling solutions, since as we have seen the kappa term can
introduce a repulsive force that prevents the big bang singularity,  there
 will very likely be a corresponding effect when we study radial collapse
of matter, and then the
analogous term, that in the homogeneous cosmology solutions prevents the
big bang singularity will in this case prevents the collapse to very high
densities. This will in turn suppress the structure formation at low
redshifts as compared to the expectations from the perturbations observed
in the CMB, thus, may be explaining the $\sigma_8$ \cite{Lambiase:2018ows}\cite{Barros:2018efl}\cite{Kazantzidis:2018rnb} - $\Omega_m$ tension.
Notice that this effect on perturbations can take place even for
constant potentials, that is without modifying the standard LCDM homogeneous background, since in the homogeneous background
the $\kappa$ term acts only in the very early universe.

Finally, another direction for research has been started by studying models of this type in the context of higher dimensional theories, where they can provide a useful framework to study the "inflation-compactication" epoch and an exit from this era to the present LCDM epoch could be further explored \cite{Benisty:2018gzx}.

\bigskip

\acknowledgments
This article is supported by COST Action CA15117 ”Cosmology and Astrophysics Network for Theoretical Advances and Training Action” (CANTATA) of the
COST (European Cooperation in Science and Technology). In addition we thank the Foundational Questions Institute FQXi for support, in particular support for our conference BASIC2018 at Stella Maris Bahamas, where part of this research was carried out.

\end{document}